\documentclass[%
reprint,
superscriptaddress,
amsmath,
amssymb,
aps,
pra,
longbibliography,
floatfix,
]{revtex4-2}

\usepackage{times}

\usepackage[colorlinks=true,linkcolor=blue,urlcolor=blue,citecolor=blue]{hyperref}
\usepackage{graphicx}
\usepackage{bm}
\usepackage{bbold}
\usepackage{dsfont}
\usepackage{xcolor}

\usepackage[normalem]{ulem}
\usepackage{scalerel}
\usepackage{tikz}
\usetikzlibrary{svg.path}
\definecolor{orcidlogocol}{HTML}{A6CE39}
\tikzset{
	orcidlogo/.pic={
		\fill[orcidlogocol] svg{M256,128c0,70.7-57.3,128-128,128C57.3,256,0,198.7,0,128C0,57.3,57.3,0,128,0C198.7,0,256,57.3,256,128z};
		\fill[white] svg{M86.3,186.2H70.9V79.1h15.4v48.4V186.2z}
		svg{M108.9,79.1h41.6c39.6,0,57,28.3,57,53.6c0,27.5-21.5,53.6-56.8,53.6h-41.8V79.1z M124.3,172.4h24.5c34.9,0,42.9-26.5,42.9-39.7c0-21.5-13.7-39.7-43.7-39.7h-23.7V172.4z}
		svg{M88.7,56.8c0,5.5-4.5,10.1-10.1,10.1c-5.6,0-10.1-4.6-10.1-10.1c0-5.6,4.5-10.1,10.1-10.1C84.2,46.7,88.7,51.3,88.7,56.8z};
	}
}
\newcommand\orcid[1]{\href{https://orcid.org/#1}{\mbox{\scalerel*{
				\begin{tikzpicture}[yscale=-1,transform shape]
					\pic{orcidlogo};
				\end{tikzpicture}
			}{R}}}}

\newcommand{\Tr}{{\rm Tr}}

\begin{document}
	\title{Robustness of random-control quantum-state tomography}
	
	\author{Jingcheng Wang\,\orcid{0000-0001-8058-1519}} 
	\affiliation{School of Physics, Hubei Key Laboratory of Gravitation and Quantum Physics, Institute for Quantum Science and Engineering, Huazhong University of Science and Technology, Wuhan 430074, China}
	\affiliation{International Joint Laboratory on Quantum Sensing and Quantum Metrology, Huazhong University of Science and Technology, Wuhan 430074, China}
	
	\author{Shaoliang Zhang\,\orcid{0000-0001-6635-044X}} 
	\affiliation{School of Physics, Hubei Key Laboratory of Gravitation and Quantum Physics, Institute for Quantum Science and Engineering, Huazhong University of Science and Technology, Wuhan 430074, China}
	\affiliation{International Joint Laboratory on Quantum Sensing and Quantum Metrology, Huazhong University of Science and Technology, Wuhan 430074, China}
	
	\author{Jianming Cai\,\orcid{0000-0002-6919-6596}} 
	\affiliation{School of Physics, Hubei Key Laboratory of Gravitation and Quantum Physics, Institute for Quantum Science and Engineering, Huazhong University of Science and Technology, Wuhan 430074, China}
	\affiliation{International Joint Laboratory on Quantum Sensing and Quantum Metrology, Huazhong University of Science and Technology, Wuhan 430074, China}
	\affiliation{State Key Laboratory of Precision Spectroscopy, East China Normal University, Shanghai, 200062, China}
	
	\author{Zhenyu Liao\,\orcid{0000-0002-1915-8559,}} 
	\affiliation{School of Electronic Information and Communications, Huazhong University of Science and Technology, Wuhan 430074, China}
	
	\author{Christian Arenz\,\orcid{0000-0001-7964-2468}} 
	\email{christian.arenz@asu.edu}
	\affiliation{School of Electrical, Computer, and Energy Engineering, Arizona State University, Tempe, Arizona 85287, USA}
	
	\author{Ralf Betzholz\,\orcid{0000-0003-2570-7267}} 
	\email{ralf\_betzholz@hust.edu.cn}
	\affiliation{School of Physics, Hubei Key Laboratory of Gravitation and Quantum Physics, Institute for Quantum Science and Engineering, Huazhong University of Science and Technology, Wuhan 430074, China}
	\affiliation{International Joint Laboratory on Quantum Sensing and Quantum Metrology, Huazhong University of Science and Technology, Wuhan 430074, China}
	
	\date{\today}

	\begin{abstract}
		In a recently demonstrated quantum-state tomography scheme [\href{https://journals.aps.org/prl/abstract/10.1103/PhysRevLett.124.010405}{Phys.~Rev.~Lett.~\textbf{124},~010405~(2020)}], a random control field is locally applied to a multipartite system to reconstruct the full quantum state of the system through single-observable measurements. Here, we analyze the robustness of such a tomography scheme against measurement errors. We characterize the sensitivity to measurement errors using the condition number of a linear system that fully describes the tomography process. Using results from random matrix theory we derive the scaling law of the logarithm of this condition number with respect to the system size when Haar-random evolutions are considered. While this expression is independent on how Haar randomness is created, we also perform numerical simulations to investigate the temporal behavior of the robustness for two specific quantum systems that are driven by a single random control field. Interestingly, we find that before the mean value of the logarithm of the condition number as a function of the driving time asymptotically approaches the value predicted for a Haar-random evolution, it reaches a plateau whose length increases with the system size. 
	\end{abstract}
	
	
	\maketitle

	\section{Introduction}
	\label{sec:intro}
	
	Using randomized measurements to reconstruct the state of a quantum system has been receiving an increasing amount of attention~\cite{Huang2020,Huang2021,Elben2020,Andreas2020}. Such randomized tomography schemes have the advantage that properties of complex quantum systems can be probed while requiring less resources than some alternative methods. Randomization is typically achieved by conjugating easily accessible observables with random unitary transformations created through randomized gates in a quantum circuit. However, the creation of these (Haar) random unitary transformations through quantum circuits typically requires access to the full system. 
	
	Among many new methods for quantum-state tomography~\cite{merkel2010random,Kyrillidis2018,Struchalin2018,Koutny2022}, methods that only require a local access to a multipartite quantum system~\cite{Peng2007,Liu2019,Xin2019,Yang2020,Betzholz2021} are particularly interesting, since full access to all constituents of a complex quantum system is rarely given in realistic settings. In particular, using random control fields to create Haar-random unitary evolutions~\cite{Banchi2017} allows for reconstructing the full quantum state when access to the system is limited. Indeed, it has recently been demonstrated~\cite{Yang2020} that randomly driving and measuring a single qubit allows for reconstructing the state of a multi-qubit system, provided the system is fully controllable, the random pulse is sufficiently long, and the time trace given by single-observable expectation-value measurements only contains negligible errors. 
	
	Here, we expand on the groundwork laid in Ref.~\cite{Yang2020} and study the robustness of this random-control tomography scheme against measurement errors. In particular, we analyze the trade-off between the accuracy in reconstructing quantum states through expectation measurements of a single observable, errors in the corresponding measurement record, and the length of the random control pulse. 
	We characterize the robustness with respect to these errors by analyzing the behavior of the condition number of a matrix that fully describes the tomography process. 
		
		We show that known results from random matrix can be utilized to develop an analytical expression for the scaling of the logarithm of this condition number with respect to the dimension of the quantum system when Haar random evolutions are considered. We go on to provide numerical evidence that the assumptions used to derive this expression are justified in the settings we consider. We remark here that the expression for the logarithm of the condition number is not only independent of the explicit form of the quantum system, namely both the system itself and the control, but also independent of how the Haar randomness of the time evolution is actually achieved, thereby making it applicable far beyond the random-control setting we focus on in the following. We then proceed by numerically investigating the temporal behavior of the logarithm of the condition number, i.e., how the asymtotic value is achieved, when the system is driven by a random control field. We focus on two case-studies, namely, a driven multi-level quantum system and an Ising spin chain where control is exerted through a single local control field. We find good agreement between the predicted value and the simulated one when the evolution time is sufficiently long. We also show an intriguing phenomenon in the convergence toward the asymptotic value for these two cases, namely plateau-like structures. While this phenomenon is not intrinsic to the local random-control tomography protocol, it is of interest for further research.

	The paper is organized as follows: In Sec.~\ref{sec:qst}, we introduce the general setup of quantum-state tomography as well as the condition number and its logarithm as a measure for the robustness against measurement errors. This is followed by a discussion of how random unitary evolutions are created via random control fields in fully controllable systems. Afterwards, we develop an analytic expression for the expected behavior of the logarithm of the condition number when Haar random unitary transformation are considered. We go on to numerically investigate, in two case studies, how Haar randomness is approached as a function of the duration of a random control field. In particular, in Sec.~\ref{sec:Nlevel}, we 
	study a $d$-level system with hopping between neighboring levels and, in Sec.~\ref{sec:spins}, a transverse-field Ising chain. In both settings, we investigate the behavior of the expected robustness as a function of the length of the random control field that is applied to the first level and the first spin, respectively. We finally draw conclusions in Sec.~\ref{sec:conclusion} and present a number of useful details in the Appendixes~\ref{app:Haar}-\ref{app:plateau}.

	\section{Robustness of random-unitary quantum-state tomography}
	\label{sec:qst}
	
	\subsection{Robustness of randomized quantum-state tomography}
	For a $d$-dimensional quantum system, a general quantum state has $d^2-1$ degrees of freedom. Let us therefore consider $\{B_n\}_{n=1}^{d^2-1}$ to be a basis for traceless Hermitian $d\times d$ matrices that is orthonormal with respect to the Hilbert-Schmidt inner product, i.e., $\Tr(B_n B_m)=\delta_{n,m}$. We collect its elements in the vector $\mathbf{B}=(B_1,\dots,B_{d^2-1})$ such that any density matrix $\rho$ of the system can be written in the form
	\begin{equation}
		\label{eq:rho}
		\rho = d^{-1}\mathds{1}_d+\mathbf{x}\cdot\mathbf{B},    
	\end{equation}
	with the $d\times d$ identity matrix $\mathds{1}_d$ and the generalized Bloch vector $\mathbf{x}$, whose components are given by $x_n=\Tr(B_n\rho)$ for $n=1,\dots,d^2-1$. 
	In order to determine $\mathbf{x}$ for an unknown quantum state, it is, in general, necessary to perform a measurement of $d^2-1$ different observables. However, these observables can also be generated from a single observable $M$ using $d^2-1$ different time evolutions, represented by the unitaries $U_n$ that effectively rotate $M$ into a set of different observables.
	
	We collect the measurement outcomes of the expectation values, up to a constant offset, in the vector $\mathbf{y}$ whose components are $y_n=\Tr(MU_n\rho U_n^\dagger )-d^{-1}\Tr(M)$. Thus, one finds the linear system of equations
	\begin{equation}
		\label{eq:matrixinvert}
		\mathcal{A}\mathbf{x}=\mathbf{y},    
	\end{equation}
	that determines the Bloch vector $\mathbf{x}$. Here, the matrix $\mathcal{A}$ connecting the Bloch vector with the measurement results has the entries $\mathcal{A}_{n,m}=\Tr(MU_nB_mU_n^\dagger)$ and fully describes the tomography process. Writing out these matrix entries in terms of the matrix entries of the observable, the basis matrices, and the unitaries gives
	\begin{equation}
		\label{eq:A2}
		\mathcal{A}_{n,m}=\sum_{j,k,p,q=1}^{d}M_{j,k}B_{p,q}^{(m)}U^{(n)}_{k,p}U^{(n)\ast}_{j,q},
	\end{equation}
	where $B_{k,l}^{(n)}$ and $U_{k,l}^{(n)}$ respectively denote the $k,l$-entries of the matrices $B_n$ and $U_n$. In the following, we will consider unitaries $U_n$ that are created uniformly random according to the Haar measure. In this case, $\mathcal A$ is almost always invertible~\cite{Yang2020}, so that the generalized Bloch vector can be obtain by inverting Eq.~\eqref{eq:matrixinvert}.  Furthermore, from the above expression, it can then be seen that for a fixed value of $m$ the entries of $\mathcal{A}$ are statistically independent random variables, because they originate from independent unitaries. However, the entries for fixed $n$ can depend on the same entries of $U_n$, making them statistically dependent.
	
	In the analysis of linear systems of equations, it is customary to describe the robustness of the solution $\mathbf{x}$ against perturbations in $\mathbf{y}$ as well as $\mathcal{A}$ with the condition number of $\mathcal{A}$. Therefore, as a figure of merit for the robustness of a faithful retrieval of $\mathbf{x}$ against errors in the measurement outcomes $\mathbf{y}$, as well as possible perturbations in the entries of $\mathcal{A}$, we use the 2-norm condition number $\kappa$~\cite{Bogdanov2010,Miranowicz2014,Miranowicz2015,Shen2016}, which is defined as
	\begin{equation}
		\kappa=||\mathcal{A}||_2  ||\mathcal{A}^{-1}||_2.   
	\end{equation}
	This expression takes the form $\kappa=s_\mathrm{max}/s_\mathrm{min}$, where $s_\mathrm{max}$ and $s_\mathrm{min}$ respectively denote the largest and smallest singular value of $\mathcal{A}$~\cite{Belsley}. For convenience, below, we will not use $\kappa$ itself but the quantity $\ln(\kappa)$ to describe the robustness of the tomography. However, before we investigate the behavior of $\ln(\kappa)$, we first describe how Haar-random evolutions can be created through randomly applied control fields in systems which are fully controllable.
	
	\subsection{Generation of Haar-random unitaries by random controls}
	\label{sec:random_control}
	We recall the random-control tomography setting in Ref.~\cite{Yang2020}, where a $d$-dimensional fully controllable quantum system~\cite{Brif2010,Schirmer2001} is driven by a single random control field. In this case, the control system is described by a time-dependent Hamiltonian of the form
	\begin{equation}
		\label{eq:Hamiltonian}
		H(t)=H_0+f(t)H_c,    
	\end{equation}
	where we refer to $H_0$ and $H_c$ as the the drift and the control Hamiltonian, respectively, and $f(t)$ represents a classical control field. 
	The unitary evolution $U(t)$ describing the dynamics of this driven quantum system is governed by the Schr\"odinger equation $\dot{U}(t)=-iH(t)U(t)$, where we have set $\hbar=1$. The system is said to be fully controllable if all unitary transformations can be achieved to arbitrary precision in finite time by appropriately shaping $f(t)$. It can be shown that this is the case iff the dynamical Lie algebra $\mathfrak{L}=\text{Lie}(iH_{0},iH_{c})$ that is formed by all nested commutators of the drift and the control Hamiltonian as well as their real linear combination span the full space, i.e., the special unitary algebra $\mathfrak{su}(d)$ for traceless Hamiltonians. For a more formal definition of full controllability we refer to~\cite{d2007introduction}.
	
	When the control system is fully controllable, a random control field yields a Haar-random evolution provided the evolution time is sufficiently long~\cite{Banchi2017}. As such, if we associate the unitaries $U_{n}$ in Eq.~\eqref{eq:A2} with different points in time of the evolution, where Haar randomness is achieved, the corresponding time trace $\mathbf{y}$ almost always contains enough information to reconstruct the generalized Bloch vector. We can consequently reconstruct the full quantum state of a complex system by randomly driving and measuring only parts of it, which has been experimentally demonstrated for an electron-nuclear spin system in Ref.~\cite{Yang2020}.

	In the subsequent analysis, if nothing else is specified, the random controls we use to generate the unitaries employ piecewise-constant control fields $f(t)$ with a segment length $\Delta t$, where the constant values of the fields in every segment are random numbers uniformly distributed on the interval $[-1,1]$ showing no temporal correlations.

	\subsection{Condition number for Haar-random unitaries}
	Let us now lay our attention on the case where the $d^2-1$ unitaries $U_n$ are generated by sufficiently long random controls, such that they can be assumed to be independently drawn from the Haar measure. The entries of $\mathcal{A}$ do not strictly follow a Gaussian distribution, thus its condition number requires further analysis. Recalling Eq.~\eqref{eq:A2} shows that the random variables $\mathcal{A}_{n,m}$ are bilinear forms of the entries of the unitaries $U_n$. Their expectation value and covariance can be calculated by integration with respect to the Haar measure (see Appendix~\ref{app:Haar} or Refs.~\cite{Hiai2000,Collins2006,Puchala2017}) which yields $\mathrm{E}[\mathcal{A}_{n,m}]=0$ and
	\begin{equation}
		\label{eq:varA}
		\mathrm{E}[\mathcal{A}_{n,m}\mathcal{A}_{n',m'}]=\frac{\delta_{n,n'}\delta_{m,m'}}{d^2-1}[\Tr(M^2)-d^{-1}\Tr(M)^2],
	\end{equation}
	and shows that they have a zero mean, are uncorrelated~\cite{Arenz2020}, and can easily be re-scaled to yield unit-variance random variables with a single normalization factor that does not depend on $n$ and $m$.
	
	The entries of $\mathcal{A}$ are not necessarily identically distributed. However, if a subset $\{B_n\}_{n=1}^D$, where $D\leq d^2-1$, is unitarily equivalent, i.e., for every $l,m\in\{1,\dots,D\}$ one can find a unitary matrix $T_{l,m}$ such that $B_l=T_{l,m} B_m T_{l,m}^\dagger$, then the random variables $\mathcal{A}_{n,1},\dots,\mathcal{A}_{n,D}$ are identically distributed for all values of $n$. This can be seen from $\mathcal{A}_{n,l}=\Tr[M(U_n T_{l,m})B_m(U_n T_{l,m})^\dagger]$ and the fact that the Haar measure is invariant under multiplication by a unitary from the right. In summary, the different columns of $\mathcal{A}$ are independent and identically distributed, whereas the entries are not independent and may follow different probability distributions in the individual rows.
	
	For $(d^2-1)\times(d^2-1)$ random matrices whose entries are independent zero-mean unit-variance random variables, it was shown that for growing system sizes the expectation value of $\ln(\kappa)$ converges to the expression~\cite{Edelman1988,Tao2009a,Tao2009b}
	\begin{equation}
		\label{eq:explogkappa}
		\mathrm{E}[\ln(\kappa)]=\ln(d^2-1)+1.537.   
	\end{equation}
	More details on this known result are provided in Appendix~\ref{app:cond}. Although, in our case, the random variables, i.e., the entries of $\mathcal{A}$, are not all statistically independent, we find that this result can still be applied to our situation. In this respect, existing works~\cite{Adamczak2011} have already shown that random matrices with statistically dependent entries can have the same asymptotic properties as shown in Eq.~\eqref{eq:explogkappa}.
	
\begin{figure}[tb]
		\includegraphics[width=\linewidth]{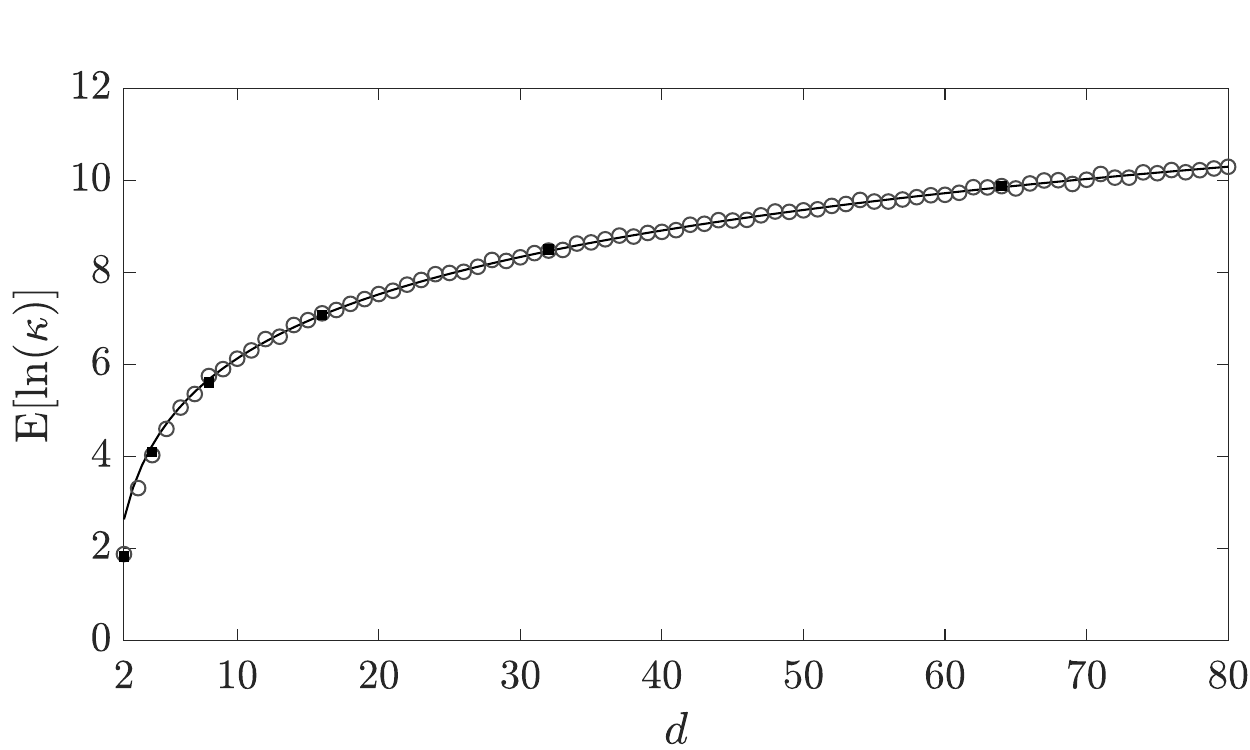}
		\caption{\label{Fig1} Dependence of the expectation value of $\ln(\kappa)$ on the dimension $d$ of the quantum system. The solid line shows the exact asymptotic result given by Eq.~\eqref{eq:explogkappa}. Open circles and filled squares represent numerical averages over 1000 realizations of $\ln(\kappa)$, based on randomly generated unitaries~\cite{Ozols2009,Cubitt2009}, for the two choices of the operator basis $\{B_n\}_{n=1}^{d^2-1}$ we use for the specific quantum systems discussed in Secs.~\ref{sec:Nlevel} and~\ref{sec:spins}, respectively.}
	\end{figure}

	In order to verify this, we have performed numerical simulations that are associated with the tomography on two specific fully-controllable quantum systems, namely a $d$-level system and a spin chain, that will be investigated in more detail in Secs.~\ref{sec:Nlevel} and~\ref{sec:spins}, respectively. The solid line in Fig.~\ref{Fig1} represents the result~\eqref{eq:explogkappa} for the dependence of $\mathrm{E}[\ln(\kappa)]$ on the dimension of the quantum system. 
	The open circles and solid squares show the numerical average over 1000 realizations of $\ln(\kappa)$, based on randomly generated unitaries~\cite{Ozols2009,Cubitt2009}, for the two specific choices of the basis $\{B_n\}_{n=1}^{d^2-1}$ we employ in the actual quantum-state-tomography setups discussed below. One sees that after initial discrepancies for very low values of the dimension, the disagreement between the numerical results and Eq.~\eqref{eq:explogkappa} becomes much less pronounced for a growing system size.

	\section{Case study 1: Multilevel system}
	\label{sec:Nlevel}
	In this section, we focus on a specific fully controllable quantum system as a first case study to verify the results for the robustness of the random-field tomography presented above.
	
	\subsection{System, control, and measurement}
	The system we investigate here is a quantum system with $d$ states denoted by $\vert n\rangle$ for $n=1,\dots, d$, with hopping terms of identical strength $h$ between adjacent levels.  Control of the system is applied in the form of an energy variation of the first level, which makes the system fully controllable~\cite{Wang2012,Burgarth2013}. The drift and the control Hamiltonian describing this control system are given by 
	\begin{gather}
		\label{eq:drift}
		H_0=h\sum_{n=1}^{d-1}(|n\rangle\langle n+1|+|n+1\rangle\langle n|),\\
		\label{eq:control}
		H_c = g|1\rangle\langle 1|.
	\end{gather}
	Such a drift Hamiltonian describes, for example, a quantum random walk~\cite{Kempe2003} with equal probabilities. 
	
	As an observable we consider the population of the first level, i.e., $M=|1\rangle\langle 1|$. As our basis, in this case, we use the generalized Gell-Mann operators~\cite{Bertlmann2008}. This basis can be split into three distinct categories, namely, the symmetric and antisymmetric matrices $B_{j,k}^{\mathrm{symm}}$ and $B_{j,k}^{\mathrm{anti}}$, respectively, for $1\leq j<k\leq d$, and the diagonal matrices $B_{l}^\mathrm{diag}$ for $1\leq l\leq d-1$. Explicitly, they are given by
	\begin{gather}
		B_{j,k}^{\mathrm{symm}}=\frac{1}{\sqrt{2}}\left(\vert k\rangle\langle j\vert+\vert j\rangle\langle k\vert\right),\\
		B_{j,k}^{\mathrm{anti}}=\frac{i}{\sqrt{2}}\left(\vert k\rangle\langle j\vert-\vert j\rangle\langle k\vert\right),\\
		B_l^{\mathrm{diag}}=\frac{1}{\sqrt{l(l+1)}}\left(\sum_{n=1}^l\vert n\rangle\langle n\vert-l\vert l+1\rangle\langle l+1\vert\right).
	\end{gather}
    By combining these three sets of matrices into a single set, with appropriate re-indexing, this becomes the basis collected in $\mathbf{B}$. The definition we employ here may differ from some definitions used elsewhere by a factor of $\sqrt{2}$, such that, in our case, the basis matrices fulfill the orthonormality relation $\Tr(B_nB_m)=\delta_{n,m}$. 
	
	With this form of the basis at hand, one can use Eq.~\eqref{eq:A2} to establish the probability distribution of the entries of $\mathcal{A}$ assuming Haar randomness of the unitaries $U_n$. The explicit form of the distribution of the $\mathcal{A}_{n,m}$ can be derived using an approximation of the entries of the unitaries by independent complex Gaussian random variables~\cite{Hiai2000,Jiang2005}, as shown in Appendix~\ref{app:case1}, even though the explicit distribution of the matrix entries of Haar-random unitaries is not readily accessible~\cite{Zyczkowski1994}.

	\subsection{Robustness}
	We will now analyze the robustness of the random-control tomography, as described by $\ln(\kappa)$, for the system introduced previously. Since we are employing random control fields $f(t)$, in order to obtain an expectation value, we take an average over a sufficiently large number of numerical realizations. We remark that the number of distinct random unitary evolutions in each numerical realization is $d^2-1$, which corresponds to the number of expectation values that have to be measured in an experiment.
	
\begin{figure}[tb]
		\flushleft\normalsize{(a)}\hspace{-2.4ex}\includegraphics[width=1\linewidth]{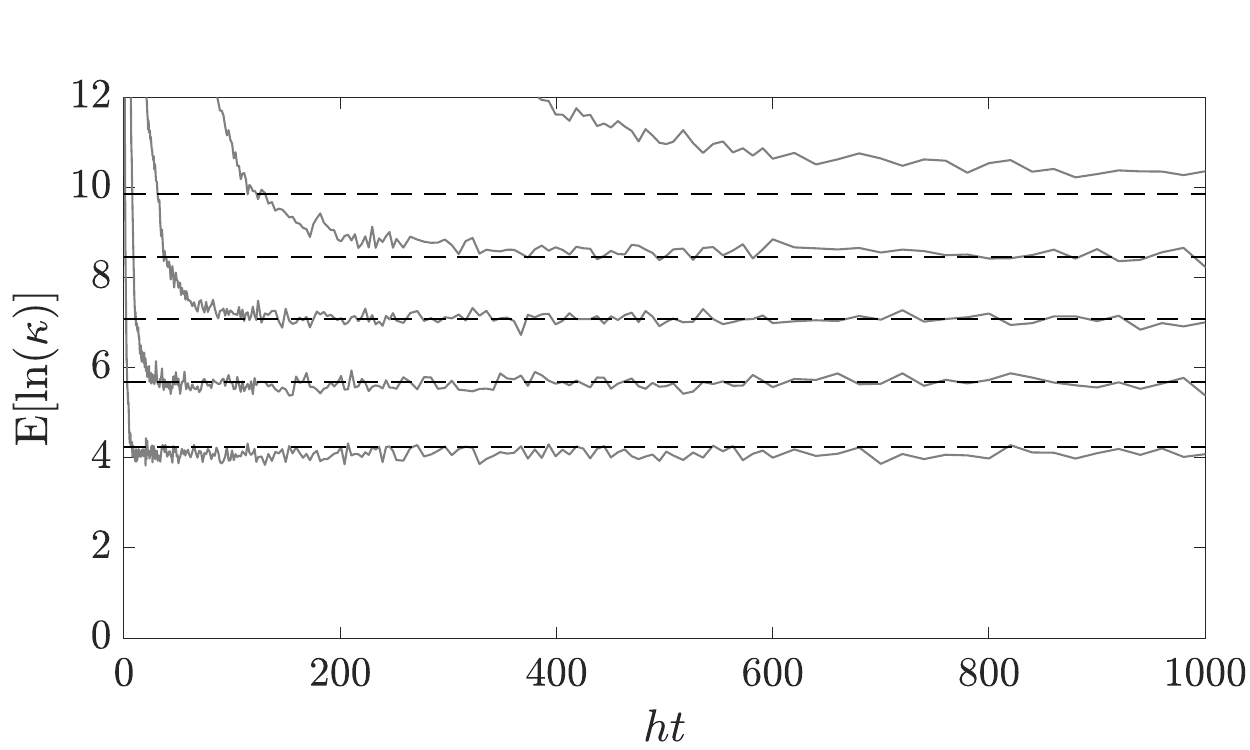}\vspace{-3ex}
		\flushleft\normalsize{(b)}\hspace{-2.4ex}\includegraphics[width=\linewidth]{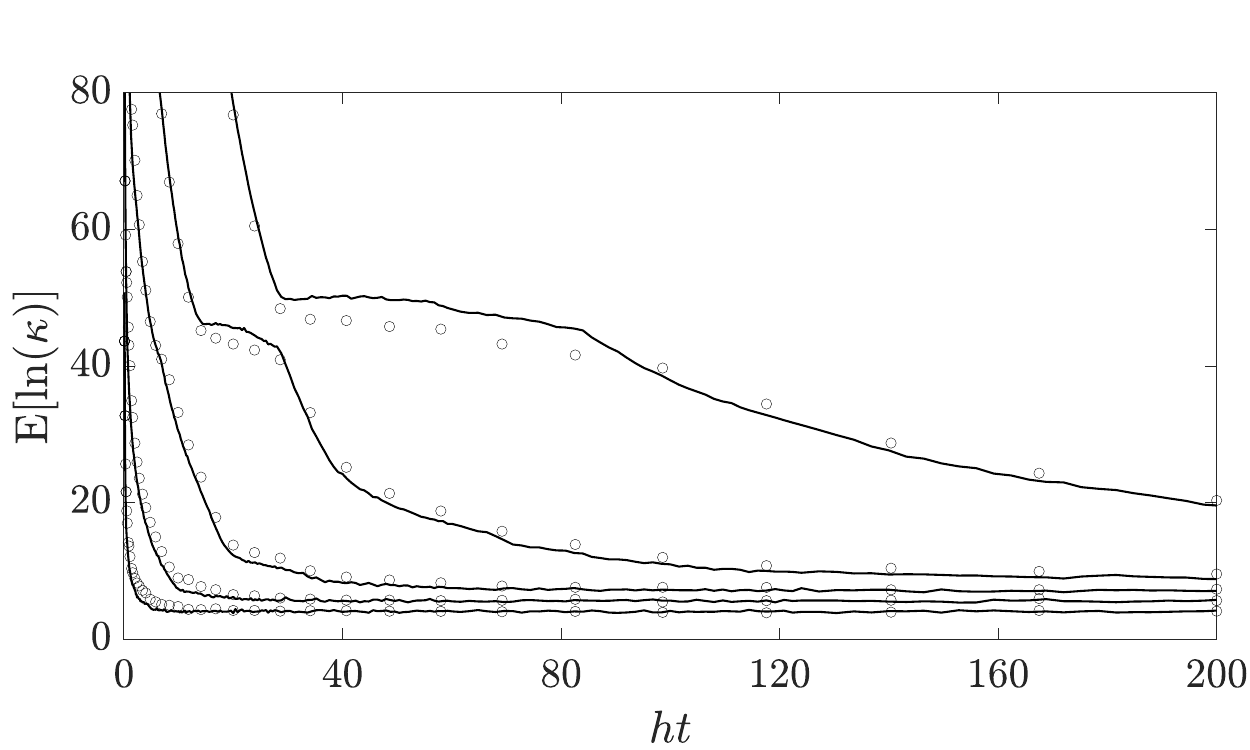}
		\caption{\label{fig:cond_Nlevel}Mean of the logarithm of the condition number of the tomography matrix for a randomly-controlled multilevel system versus the duration $t$ of a single random unitary evolution. The result is averaged over 100 realizations with the parameters $g/h=10$, $h\Delta t=0.1$, $K=20$, and $\Omega=g$, with each realization using $d^2-1$ distinct random unitary evolutions: (a) Long-time behavior for piecewise-constant control for $d=4,8,16,32,64$ (bottom to top). 
		The dashed horizontal lines indicate the value of $\mathrm{E}[\ln(\kappa)]$ for Haar-random unitaries, which are given by Eq.~\eqref{eq:explogkappa} and shown in Fig.~\ref{Fig1}. (b) Short-time behavior, where solid lines and circles correspond to piecewise-constant and truncated-Fourier controls, respectively.}
	\end{figure}

	Fig.~\ref{fig:cond_Nlevel} shows the time evolution of $\mathrm{E}[\ln(\kappa)]$ calculated using $100$ realizations of the random pulses with a pulse-segment length fulfilling $h\Delta t=0.1$ and a control-field amplitude $g/h=10$. 
	The long-time behavior of the average of $\ln(\kappa)$ is shown in Fig.~\ref{fig:cond_Nlevel}(a) for different dimensions of the system, namely $d=4,8,16,32,64$. There, we observe convergence toward the value expected for Haar random unitaries (shown as dashed horizontal lines), which are given by Eq.~\eqref{eq:explogkappa} and shown in Fig.~\ref{Fig1} as a function of $d$. The time $t$ shown in Fig.~\ref{fig:cond_Nlevel} denotes the duration of a single random unitary evolution.
	
	Having a closer look at shorter times reveals more subtleties in the convergence behavior. Here, as a second example for a random control, we also employ a truncated Fourier series~\footnote{Such control fields may be advantageous for experimental implementations due to their continuous nature.} of the form $f(t)=\sum_{k=1}^KF_k\cos(\omega_kt+\varphi_k)$, with uniformly distributed parameters: $F_k$ fulfilling $\sum_{k=1}^KF_k=1$, $\omega_k\in[0,\Omega]$, and $\varphi_k\in[0,2\pi]$. In Fig.~\ref{fig:cond_Nlevel}(b), one sees that after a rapid initial drop for very short times,  a plateau appears over which the slope is significantly lower than during the initial decrease. This behavior is present for both kinds of random controls we used (circles for truncated Fourier series and solid lines for piecewise constant). After this stage, there appears another more rapid drop. During the convergence to the asymptotic value, more of such plateaus can be found, although not as pronounced as the first one. We found that the length of the first plateau increases for a growing number $d$ of levels and the time $t_p$ until it is reached can be well approximated by the linear dependence $ht_p\propto d$ on the dimension. 
	
	More details on this phenomenon are given in App.~\ref{app:plateau}, including the observation that this behavior is special to our case studies and is not necessarily found in other controlled quantum systems. We believe that it is closely related to the geometry of the coupling graph of the system and how the control spreads through the system.

	\section{Case study 2: Spin system}
	\label{sec:spins}
	As a second example, we consider a fully controllable system comprised of $N$ spins, i.e., of the dimension $d=2^N$. 
	
	\subsection{System, control, and measurement}
	In particular, we investigate an Ising chain with a transverse field in the $x$-$z$ plane and a control that is applied only to a single spin at the edge of the chain. Therefore, we consider the drift and control Hamiltonians given by
	\begin{gather}
		\label{eq:spin-systems1}
		H_0=h\left[\sum_{n=1}^{N-1}\sigma_n^z\sigma_{n+1}^z+\sum_{n=1}^N(\sigma_n^x+\sigma_n^z)\right],\\
		\label{eq:spin-systems2}
		H_c=g\sigma_1^x,
	\end{gather}
	where the $\sigma_n^\alpha$ denote the Pauli matrices of the $n$th spin, for $\alpha=x,y,z$. Here, we have, for the sake of convenience, set the nearest-neighbor interaction equal to the global field strength $h$. However, independently of this assumption we show in Appendix~\ref{app:full_ctrl} that the control system given by the pair Eqs.~\eqref{eq:spin-systems1} and~\eqref{eq:spin-systems2} is fully controllable. We choose $M$ to be a single-spin observable, namely $M=\sigma_1^z$. For this special case of the Hilbert-space dimension one can use the Pauli matrices to construct the basis whose elements are given by the products
	\begin{equation}
		B_n=2^{-N/2}\prod_{m=1}^N \sigma_m^{k_m},
	\end{equation}
	where every value of $n$ corresponds to one of the $4^N-1$ tupels $(k_1,\dots,k_N)\in\{0,x,y,z\}^N \setminus \{0,0,0,0\}$, with $\sigma_j^0=\mathds{1}_2$. In this case, all basis matrices are unitarily equivalent, which means that for Haar random unitaries all entries of $\mathcal{A}$ are identically distributed. In Appendix~\ref{app:case2}, it is furthermore shown that, in the case of Haar-random unitaries, a good approximation for their probability distribution is given by a normal distribution.

	\subsection{Robustness}
\begin{figure}[tb]
		\flushleft\normalsize{(a)}\hspace{-2.4ex}\includegraphics[width=1\linewidth]{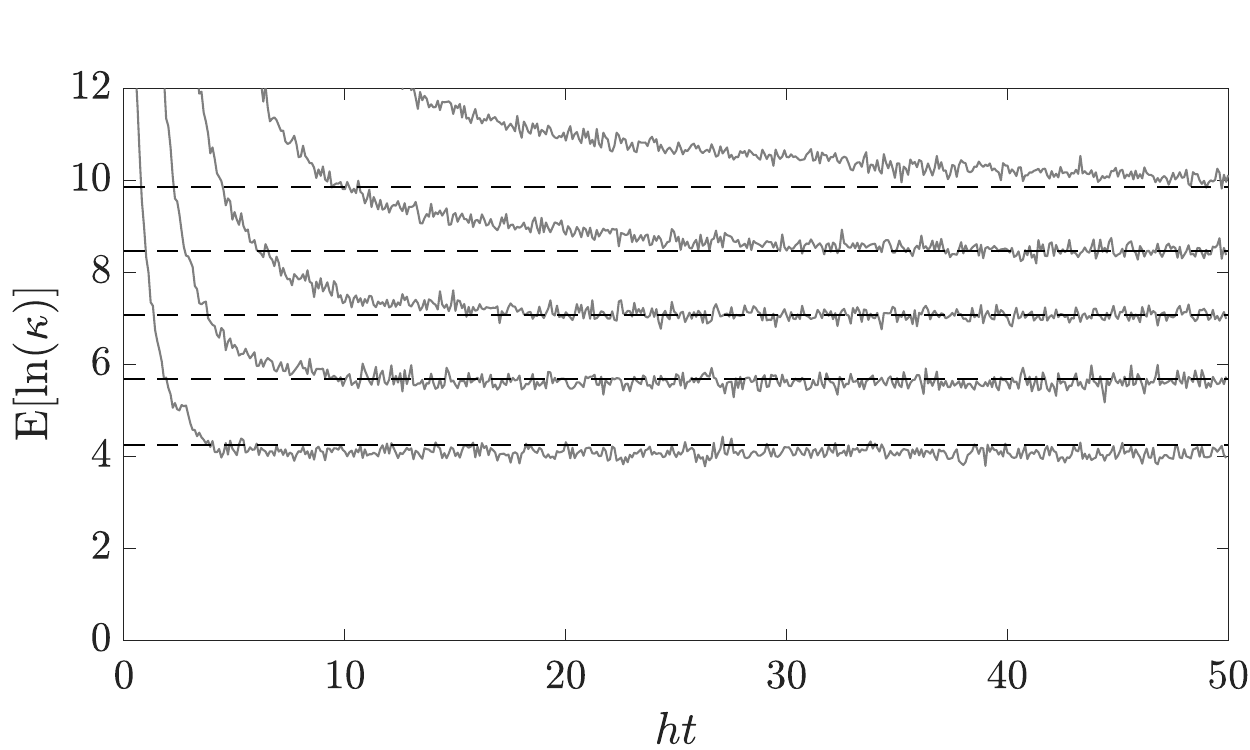}\vspace{-3ex}
		\flushleft\normalsize{(b)}\hspace{-2.4ex}\includegraphics[width=1\linewidth]{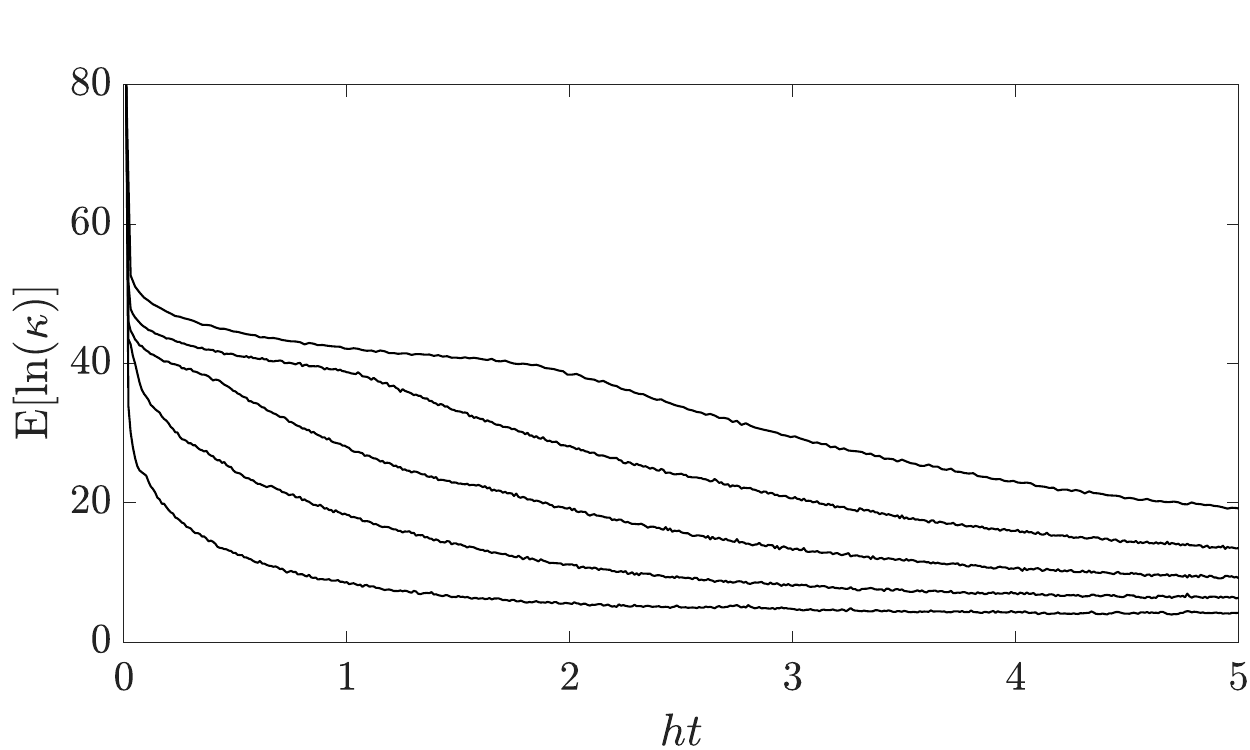}\vspace{-3ex}
		\caption{\label{fig:spincond}The mean of the logarithm of the condition number, $\ln(\kappa)$, of the tomography matrix for a randomly-controlled Ising spin chain in dependence of the duration $t$ of a single random unitary evolution. The result is averaged over 100 realizations with the parameters $g/h=10$ and $h\Delta t=0.01$. Each realization uses $4^N-1$ distinct random unitary evolutions: (a) Long-time behavior for $N=2,3,4,5,6$ (bottom to top). The dashed horizontal lines show the value of $\mathrm{E}[\ln(\kappa)]$ for Haar-random unitaries, which are given by Eq.~\eqref{eq:explogkappa} and shown as solid squares in Fig.~\ref{Fig1}. (b) Short-time behavior for the same values of $N$.}
	\end{figure}

	In Fig.~\ref{fig:spincond}, we present the results of the robustness for this spin system in the same fashion as we have for the multilevel system. That is, we again show an average of $\ln(\kappa)$ over 100 realizations, with $4^N-1$ distinct random unitary evolutions for each realization, although shown here only for piecewise-constant random pulses. We used the parameters $g/h=10$ and $h\Delta t=0.01$ for $N=2,3,4,5,6$ spins, which corresponds to the same dimensions as for the multilevel system. Here, the time $t$ shown in Fig.~\ref{fig:spincond} denotes the duration of a single random unitary evolution as well.
	
	In Fig.~\ref{fig:spincond}(a) we again see convergence toward the asymptotic value expected for Haar random unitaries (shown as dashed horizontal lines), which are given by Eq.~\eqref{eq:explogkappa} and represented as solid squares in Fig.~\ref{Fig1}. The short-time behavior is shown in Fig.~\ref{fig:spincond}(b). Similar to the previous case study, the initial drop is rapid and is followed by a plateau over which $\mathrm{E}[\ln(\kappa)]$ changes much slower. Also here, this plateau becomes more pronounced and longer for higher dimensions.

	\section{Conclusions}
	\label{sec:conclusion}
	We have established an expression for the robustness against measurement errors of randomized quantum-state tomography. This robustness is quantified by the condition number $\kappa$ of a matrix that fully describes the tomography process. We have derived an analytical expression for the mean of its logarithm when randomization is achieved through applying Haar-random unitary to a given observable. In particular, we have shown that, up to a constant, the dependence on the system dimension $d$ of the mean value of $\ln(\kappa)$ is given by $\ln(d^2-1)$. This result constitutes a general property of quantum-state tomography based on Haar-random unitaries since it depends neither on the specific kind of quantum system under investigation nor on the actual method that is employed for the generation of the Haar-random unitaries. 
	
	We went on to numerically investigate the asymptotic behavior in more detail for the case when Haar randomness is achieved through the application of random control fields. Here, we have studied the robustness of such a random-control quantum-state tomography scheme for two distinct controlled quantum systems; a $d$-level system with hopping between neighboring levels and a transverse-field Ising chain. In both cases, the system is fully controllable through a single control field that is applied locally, which we used together with a local observable measurement to achieve full quantum-state reconstruction~\cite{Yang2020}. The random drive asymptotically leads to a Haar-random time evolution, such that the robustness converges to the above result as the control time tends to infinity. However, in the two quantum systems we investigated as case studies, we have found that while a finite time evolution is sufficient to achieve a satisfactory convergence toward the asymptotic value, the numerical simulations suggest that before the asymptotic value is reached, a plateau appears whose length increases with the system size. It would be interesting to relate this behavior to the cutoff phenomenon in classical Markov chains~\cite{diaconis1996cutoff}, which will be subject of future studies.

	\begin{acknowledgments}
		We thank S. B. J\"ager, J. M. Torres, and J. H. Zhang for helpful discussions and G. Morigi for the kind opportunity to use the computational resources of the Theoretical Quantum Physics Group at Saarland University during the early stages of this work. J.~W., S.~Z., and J.~C. acknowledge support from the National Natural Science Foundation of China (Grants No.~12161141011, No.~11874024, No.~11690032, and No.~12174138), the National Key R\&D Program of China (Grant No.~2018YFA0306600), the Fundamental Research Funds for the Central Universities, the Open Project Program of the Shanghai Key Laboratory of Magnetic Resonance, and the Interdisciplinary Program of the Wuhan National High Magnetic Field Center (Grant No.~WHMFC202106). Z.~L. is supported by the National Natural Science Foundation of China (Grant No.~62206101 and No.~12141107) and the Fundamental Research Funds for the Central Universities of China (2021XXJS110). C. A. acknowledges support from the National Science Foundation (Grant No. 2231328). R.~B. is supported by startup funding of the Huazhong University of Science and Technology. Part of the computation was completed on the HPC Platform of Huazhong University of Science and Technology.
	\end{acknowledgments}

	\appendix
	
	\section{Integration with respect to the Haar measure}
	\label{app:Haar}
	For Haar-random unitaries $U$ the vanishing expectation value and the covariance~\eqref{eq:varA} of the entries of $\mathcal{A}$ were presented in the main text. These results can be easily established by integration over the unitary group $\mathrm{U}(d)$ with respect to the Haar probability measure $\mu$~\cite{Hiai2000,Collins2006,Puchala2017}. In detail, one can employ
	\begin{equation}
		\label{eq:intU1}
		\int_{\mathrm{U}(d)}d\mu(U)\, U_{i,j}U^\ast_{k,l}=d^{-1}\delta_{i,k}\delta_{j,l}
	\end{equation}
	and
	\begin{align}
		\int_{\mathrm{U}(d)}d\mu(U)\,& U_{i,j}U_{k,l}U^\ast_{i',j'}U^\ast_{k',l'}=\nonumber\\
		&\frac{\delta_{i,i'}\delta_{k,k'}\delta_{j,j'}\delta_{l,l'}+\delta_{i,k'}\delta_{k,i'}\delta_{j,l'}\delta_{l,j'}}{d^2-1}\nonumber\\
		\label{eq:intU2}
		&-\frac{\delta_{i,i'}\delta_{k,k'}\delta_{j,l'}\delta_{l,j'}+\delta_{i,k'}\delta_{k,i'}\delta_{j,j'}\delta_{l,l'}}{d(d^2-1)}.
	\end{align}
	The expectation value of a function $p(U)$ can then be calculated according to $\mathrm{E}[p(U)]=\int_{\mathrm{U}(d)}d\mu(U)\,p(U)$. Applied to Eq.~\eqref{eq:A2} this leads to $\mathrm{E}[\mathcal{A}_{n,m}]=0$ and Eq.~\eqref{eq:varA} from the main text, where we used the fact that the basis matrices are traceless and orthonormal.
	
	\section{Probability density of the condition number}
	\label{app:cond}
	In the seminal paper Ref.~\cite{Edelman1988}, it was shown that for $n\times n$ random matrices whose entries are independent real standard normal variables the quantity $\kappa/n$ converges in distribution to a random variable with the probability density 
	\begin{equation}
		\label{eq:cond_density}
		f_{\kappa/n}(x)=\frac{2x+4}{x^3}e^{-2/x-2/x^2}.
	\end{equation}
	While the integrals to compute moments of $\kappa/n$ do not exist, the ones to calculate the moments of $\ln(\kappa/n)$ can be evaluated explicitly. For the expectation value $\mathrm{E}[\ln(\kappa/n)]$ one can use the integral
	\begin{align}
		&\int_0^\infty d x\,\frac{2x+4}{x^3}e^{-2/x-2/x^2}\ln(x)=\nonumber\\
		&\frac{1}{2}\left[\gamma+\frac{\ln(4)}{2}+\sqrt{2\pi e}+\sqrt{\frac{\pi e}{2}}M\left(0,\frac{1}{2},-\frac{1}{2}\right)\right.\nonumber\\
		&\left.-\sqrt{\frac{\pi e}{2}}M\left(0,\frac{3}{2},-\frac{1}{2}\right)-M\left(1,\frac{1}{2},\frac{1}{2}\right)+M\left(1,\frac{3}{2},\frac{1}{2}\right)\right]
	\end{align}
	with the Euler–Mascheroni constant $\gamma$ and Kummer's function $M(a,b,z)$~\cite{Abramowitz}. The numerical value of this is given by $1.537$ and applied to the case of the main text, namely $n=d^2-1$, one readily obtains Eq.~\eqref{eq:explogkappa}.
	
	The above result was further generalized, in Refs.~\cite{Tao2009a,Tao2009b}, to random matrices with independent zero-mean unit-variance entries with an arbitrary, not necessarily identical, distribution.

	\section{Distribution of the difference of two Gamma-distributed random variables}
	\label{app:diffGamma}
	For the setup discussed in Sec.~\ref{sec:Nlevel}, the probability distribution of the entries of $\mathcal{A}$ can be derived explicitly by approximating the entries of the Haar-random unitaries with independent complex Gaussian random variables. Under this approximation, one comes across the difference of two Gamma-distributed random variables, as we will see below. Let us therefore derive the general form of the probability density of such differences in detail. As a first step, we employ the fact that if $X$ and $Y$ have the probability densities $f_X(x)$ and $f_Y(y)$, respectively, then $Z=X-Y$ has the probability density
	\begin{equation}
		f_Z(z)=\int_{-\infty}^\infty d x\, f_X(x)f_{Y}(x-z),
	\end{equation}
	where we have used that the variable $-Y$ has the density $f_{-Y}(y)=f_Y(-y)$. We now consider the case where $X\sim\mathrm{Gamma}(\alpha_1,\beta_1)$ and $Y\sim\mathrm{Gamma}(\alpha_2,\beta_2)$ and remind ourselves that the probability density of a $\mathrm{Gamma}(\alpha,\beta)$-distributed random variable is $\beta^\alpha x^{\alpha-1}e^{-\beta x}/\Gamma(\alpha)$ for $x\geq0$ and zero otherwise. The integral above can be evaluated using, for example, 3.383 (4) in Ref.~\cite{Gradshtyn}, yielding the final expression~\cite{Klar2015}
	\begin{align}
		\label{eq:whittaker}
		f_{Z}(z)&=\frac{\beta_1^{\alpha_1}\beta_2^{\alpha_2}  |z|^{\frac{\alpha_1+\alpha_2}{2}-1}   e^{\frac{\beta_2-\beta_1}{2}z} }{(\beta_1+\beta_2)^{\frac{\alpha_1+\alpha_2}{2}}}\nonumber\\
		\times&
		\begin{cases}
			\displaystyle
			\frac{1}{\Gamma(\alpha_1)}   W_{\frac{\alpha_1-\alpha_2}{2},\frac{1-\alpha_1-\alpha_2}{2}}\bm{(}(\beta_1+\beta_2)z\bm{)},  & z\geq 0 \\
			\displaystyle
			\frac{1}{\Gamma(\alpha_2)}    W_{\frac{\alpha_2-\alpha_1}{2},\frac{1-\alpha_1-\alpha_2}{2}}\bm{(}(\beta_1+\beta_2)|z|\bm{)}, & z<0,
		\end{cases}
	\end{align}
	with the Whittaker function $W_{\kappa,\mu}(z)$~\cite{Abramowitz,NIST}.

	\section{Approximate distribution of the matrix entries in case study 1}
	\label{app:case1}
	Every projector $P$ can be transformed according to $RPR^\dagger=|1\rangle\langle 1|$ with a unitary matrix $R$. Thereby, the invariance of the Haar measure ensures that the same results we derive below for the simple choice $M=|1\rangle\langle 1|$ apply to all observables that are projectors. In the Gell-Mann basis, one finds the unitarily equivalent subset $\{B_{j,k}^{\mathrm{sym}},B_{j,k}^{\mathrm{anti}}, B_{1}^{\mathrm{diag}}\}_{1\leq j<k\leq d}$. This means the $D=d^2-d+1$ the corresponding variables are thereby identically distributed. Let us, for the sake of convenience, choose them to be distributed as the one corresponding to $B^\mathrm{diag}_1$. The remaining $d-2$ variables then correspond to $B^\mathrm{diag}_n$, for $n\in\{2,\dots,d-1\}$. Overall, this means it is sufficient to analyze the $d-1$ random variables
	\begin{equation}
		\label{eq:nu1}
		V_n=\frac{1}{2\sqrt{n(n+1)}}\left(\sum_{k=1}^n|U_{1,k}|^2
		- n|U_{1,n+1}|^2\right).
	\end{equation}
	
	Since it is far from straightforward to obtain the explicit probability distribution of the individual matrix entries of Haar random unitaries (see, e.g., Ref.~\cite{Zyczkowski1994}), it would be favorable to bring the above expression into a form that is easier to handle. To achieve this, we use the fact that for $d\times d$ Haar random matrices $U$ the matrix $\sqrt{d}U$ converges in distribution to a standard complex Gaussian matrix (see Lemma 4.2.4 in Ref.~\cite{Hiai2000} or Theorem 6 of Ref.~\cite{Jiang2005}). We can therefore approximate the entries $\{U_{j,k}\}_{j,k=1}^{d}$ by complex normal variables.
	
	Applying this approximation to the re-scaled variables $\nu_n=dV_n$ yields $\nu_n=X_n-Y_n$, where $X_n$ and $Y_n$ are respectively distributed according to $\mathrm{Gamma}\bm{(}n,\sqrt{n(n+1)}\bm{)}$ and $\mathrm{Gamma}\bm{(}1,\sqrt{(n+1)/n}\bm{)}$. Therefore, we can directly apply Eq.~\eqref{eq:whittaker}. We find $\beta_1\pm\beta_2=\sqrt{(n+1)/n}(n\pm 1)$ and can use $W_{\kappa,\mu}(z)=W_{\kappa,-\mu}(z)$ as well as the properties
	\begin{gather}
		W_{\mu-\frac{1}{2},\mu}(z)=e^{\frac{z}{2}}z^{\frac{1}{2}-\mu}\Gamma(2\mu,z),\\
		W_{\kappa,\kappa-\frac{1}{2}}(z)=e^{-\frac{z}{2}}z^{\kappa},
	\end{gather}
	with the incomplete Gamma function $\Gamma(k,z)$~\cite{NIST}. By identifying $\mu=n/2$ and $\kappa=1/2-n/2$, this simplifies the probability density to
	\begin{align}
		f_{\nu_n}(x)=&\left(\frac{n}{n+1}\right)^{n-\frac{1}{2}} e^{\sqrt{\frac{n+1}{n}} x}\nonumber\\ 
		&\times\begin{cases}
			\dfrac{\Gamma\bm{(}n,(n+1)^{\frac{3}{2}} n^{-\frac{1}{2}}x\bm{)}}{(n-1) !}, & x \geq 0 \\ 1, & x<0.
		\end{cases}
	\end{align}
	
	For $n=1$ we can use $\Gamma(1,z)=\exp(-z)$ to find the exponential distribution $f_{1}(x)=\exp(-\sqrt{2}|x|)/\sqrt{2}$.
	When $n$ tends to infinity, on the other hand, we find $\lim_{n\to\infty} f_n(x)=\exp(x-1)\Theta(1-x)$, with the Heaviside step function $\Theta$~\footnote{To obtain this limiting distribution, one can use the limit $\lim _{n \rightarrow \infty} \Gamma\bm{(}n,(n+1)^{\frac{3}{2}} n^{-\frac{1}{2}}x\bm{)}/\Gamma(n)=\Theta(1-x)$, which follows from 6.5.34 in Ref.~\cite{Abramowitz}}.

	\section{Proof of full controllability in case study 2}
	\label{app:full_ctrl}
	Here, we show that the control system given by
	\begin{gather}
		\label{eq:controlsystem1}
		H_0=\sum_{n=1}^{N-1}\sigma_n^z\sigma_{n+1}^z+\sum_{n=1}^N(\sigma_n^x+\sigma_n^z),\\
		\label{eq:controlsystem2}
		H_c=\sigma_1^x,
	\end{gather}
	is fully controllable. For convenience, in the proof, we use a rescaled and dimensionless version of the spin-chain Hamiltonians of the main text, viz. Eqs.~\eqref{eq:spin-systems1} and~\eqref{eq:spin-systems2}. We show that the dynamical Lie algebra $\mathfrak{L}=\text{Lie}(iH_{c},iH_{0})$ that is formed by all nested commutators of the drift $iH_{0}$ and the control $iH_{c}$ as well as their real linear combinations spans the special unitary algebra $\mathfrak{su}(2^{N})$. Principally, the proof can be summarized as follows. We first show that $\mathfrak{su}_{1}(2)=\mathfrak{su}(2)\otimes\mathds{1}_{2^{N-1}}$ of the first spin of the chain is contained in $\mathfrak{L}$. It inductively follows that then also $\mathfrak{su}_{j}(2)$, $j=2,\dots, N$, of all other spins is contained in $\mathfrak L$. Recalling results from Ref.~\cite{Zeier2011} we conclude that the control system \eqref{eq:controlsystem1}~-~\eqref{eq:controlsystem2} is thereby fully controllable. 
	
	Let us now construct this proof explicitly. The commutator of the drift and the control Hamiltonian gives the Lie-algebra element $a_{1}=i(\sigma_{1}^{y}+\sigma_{1}^{y}\sigma_{2}^{z})$. By commuting $a_{1}$ with the control Hamiltonian anew, one obtains the Lie-algebra element $a_{2}=i(\sigma_{1}^{z}+\sigma_{1}^{z}\sigma_{2}^{z})$, which can then be used to create, together with $iH_{c}$ and $iH_{0}$, the Lie-algebra element 
	\begin{align}
		a_{3}=i\left[\sum_{n=2}^{N-1}\sigma_n^z\sigma_{n+1}^z+\sum_{n=2}^N(\sigma_n^x+\sigma_n^z)\right]
	\end{align}
	via linear combination. Commuting $a_{3}$ twice with $a_{2}$ gives the Lie-algebra element $a_{4}=i\sigma_{1}^{y}\sigma_{2}^{y}$, from which one retrieves $a_{5}=i\sigma_{1}^{x}\sigma_{2}^{y}$ through commutation with $a_{2}$. The commutator of $a_{4}$ and $a_{5}$ finally gives $i\sigma_{1}^{z}$, which implies that $\mathfrak{su}_{1}(2)\in\mathfrak{L}$. We proceed by noting that the commutator of $a_{1}$ and $a_{2}$ gives, via linear combination with $iH_{c}$, the element $a_{6}=i\sigma_{1}^{x}\sigma_{2}^{z}$, from which one, in turn, obtains through commutation with $a_{5}$ the Lie-algebra element $a_{7}=i\sigma_{2}^{x}$. We can thus repeat the procedure above, this time starting with the pair $a_{3}$ and $a_{7}$, to establish $\mathfrak{su}_{2}(2)\in\mathfrak{L}$. It inductively follows that then also $\mathfrak{su}_{j}(2)\in \mathfrak{L}$, $j=1,\dots,N$. The proof can then be concluded by one final step. Since the coupling graph describing the spin system, namely a chain, is connected, it follows from Theorem 1 in Ref.~\cite{Zeier2011} that the control system described by Eqs.~\eqref{eq:controlsystem1} and~\eqref{eq:controlsystem2} is fully controllable.

	\section{Approximate distribution of the matrix entries in case study 2}
	\label{app:case2}
	In Sec.~\ref{sec:spins}, we have already established that for the setup discussed there all basis matrices are unitarily equivalent. For the sake of simplicity, we will therefore focus on the simple basis matrix $\sigma_1^z$. The random variable~\eqref{eq:A2} then takes the form
	\begin{align}
		V=2^{-N/2} \sum_{j,k=1}^{2^{N-1}}(& |U_{j,k}|^2+|U_{j+2^{N-1},k+2^{N-1}}|^2\nonumber\\
		&-|U_{j+2^{N-1},k}|^2-|U_{j,k+2^{N-1}}|^2 ).
	\end{align}
	
	Here, we can again approximate $2^{N/2}U$ by a standard complex Gaussian random matrix. The normalized random variable $\nu=2^{N/2}V$ thereby has the form $\nu=X-Y$, where $X$ and $Y$ are two independent $\mathrm{Gamma}(4^N/2,2^N)$-distributed random variables. To calculate the probability density of $\nu$ we can again use Eq.~\eqref{eq:whittaker} with $\alpha_1=\alpha_2=4^N/2$ and $\beta_1=\beta_2=2^N$ as well as the identity $W_{0,\mu}(2z)=\sqrt{2z/\pi}K_\mu(z)$ (cp. 13.18.9 of Ref.~\cite{NIST}), with the modified Bessel function of the second kind $K_\mu(z)$, in order to find the Bessel distribution~\cite{Johnson}
	\begin{equation}
		f_\nu(x)=\frac{2^N}{\sqrt{\pi}\Gamma(4^N/2)}(2^{N}|x|/2)^{\frac{4^N}{2}-\frac{1}{2}}K_{\frac{4^N}{2}-\frac{1}{2}}(2^N|x|).
	\end{equation}
	The limiting distribution of this is given by nothing but the standard normal distribution, viz., $\lim_{N\to\infty}f_\nu(x)=\exp(-x^2/2)/\sqrt{2\pi}$~\footnote{This can be proven, e.g., by showing the equality of their series expansions~\cite{Watson} using $ \lim_{n\to\infty}n^k\Gamma(n-k)/\Gamma(n)=1$ in every term.}. Already for $N=3$ the probability density is almost indistinguishable from the normal distribution and for $N>3$ the shape does not visibly change anymore.

	\section{Analysis of the plateaus in case study 1}
	\label{app:plateau}
	Below, we present a brief analysis of the plateaus that arise during the convergence toward the asymptotic value of $\mathrm{E}[\ln(\kappa)]$ in our case study 1. Before we do so, we start by making an important observation, namely, that the emergence of these plateaus has its origins most likely in the special structure of the controlled quantum systems at hand. This is supported by numerical evidence where randomly choosing a pair of drift and control, that form a fully controllable system, does not produce a similar behavior.  
		
	We therefore conjecture that the observed phenomenon originates in the geometry of the underlying quantum system, namely, in the structure of its coupling graph. For both our case studies, this is a chain-like graph with the control applied to one of the ends. However, a more definitive explanation lies outside the scope of this work and will be shown elsewhere. Nevertheless, we present some preliminary perspective.
	
	 In our random-field tomography setup, the unitaries are initially all the identity matrix $\mathds{1}_d$ and become Haar random only asymptotically. In the two previous Appendixes~\ref{app:case1} and~\ref{app:case2}, we have already made use of the fact that for a $d\times d$ Haar random unitary $U$ the matrix entries of $\sqrt{d}U$ can be approximated by standard complex normal variables. Let us consider a single entry of the unitary, e.g., the very first entry $U_{1,1}$. The mean of the magnitude of this entry starts at the initial value one and converges toward the value $\mathrm{E}[|U_{1,1}|]$. Since $\sqrt{d}U_{1,1}$ can be approximated by a standard complex normal variable, the magnitude $\sqrt{d}|U_{1,1}|$ is Rayleigh distributed with a scale parameter of $1/\sqrt{2}$. Thereby, one can establish the asymptotic value $\mathrm{E}[|U_{1,1}|]=\sqrt{\pi/4d}$.
	 
\begin{figure}[tb]
	 	\flushleft\normalsize{(a)}\hspace{-2.4ex}\includegraphics[width=1\linewidth]{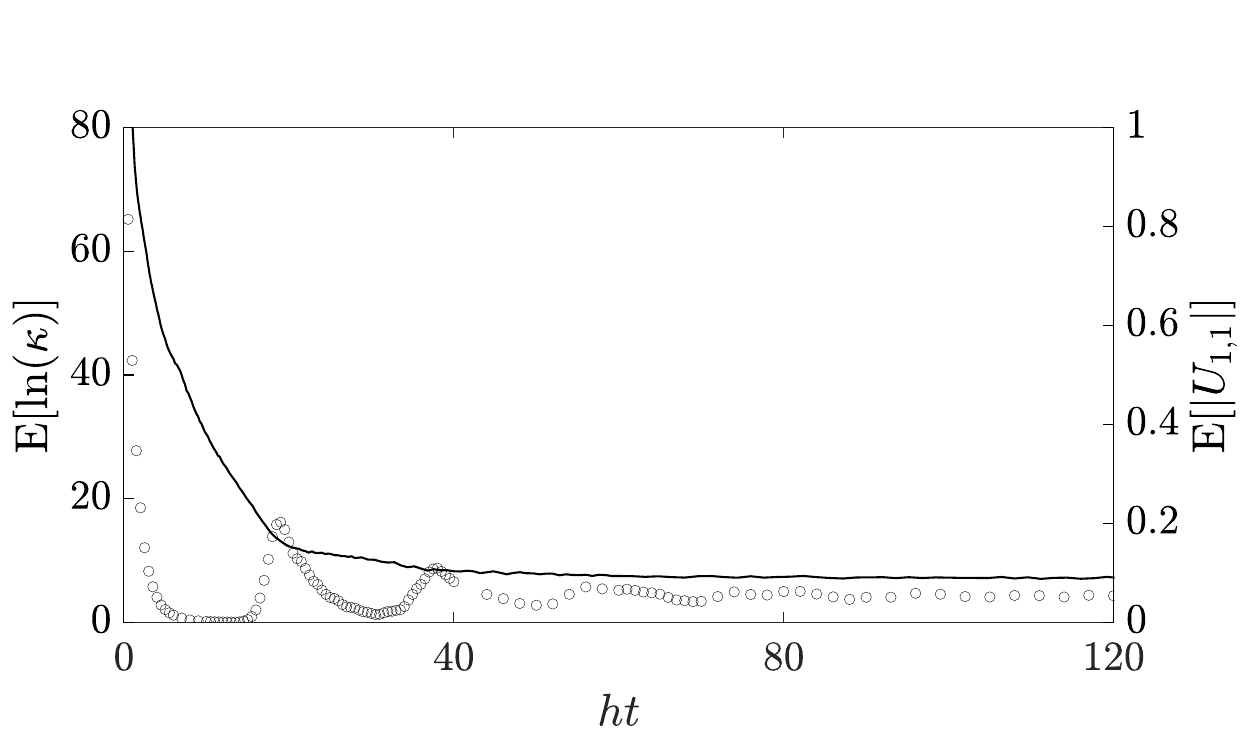}\vspace{-3ex}
	 	\flushleft\normalsize{(b)}\hspace{-2.4ex}\includegraphics[width=\linewidth]{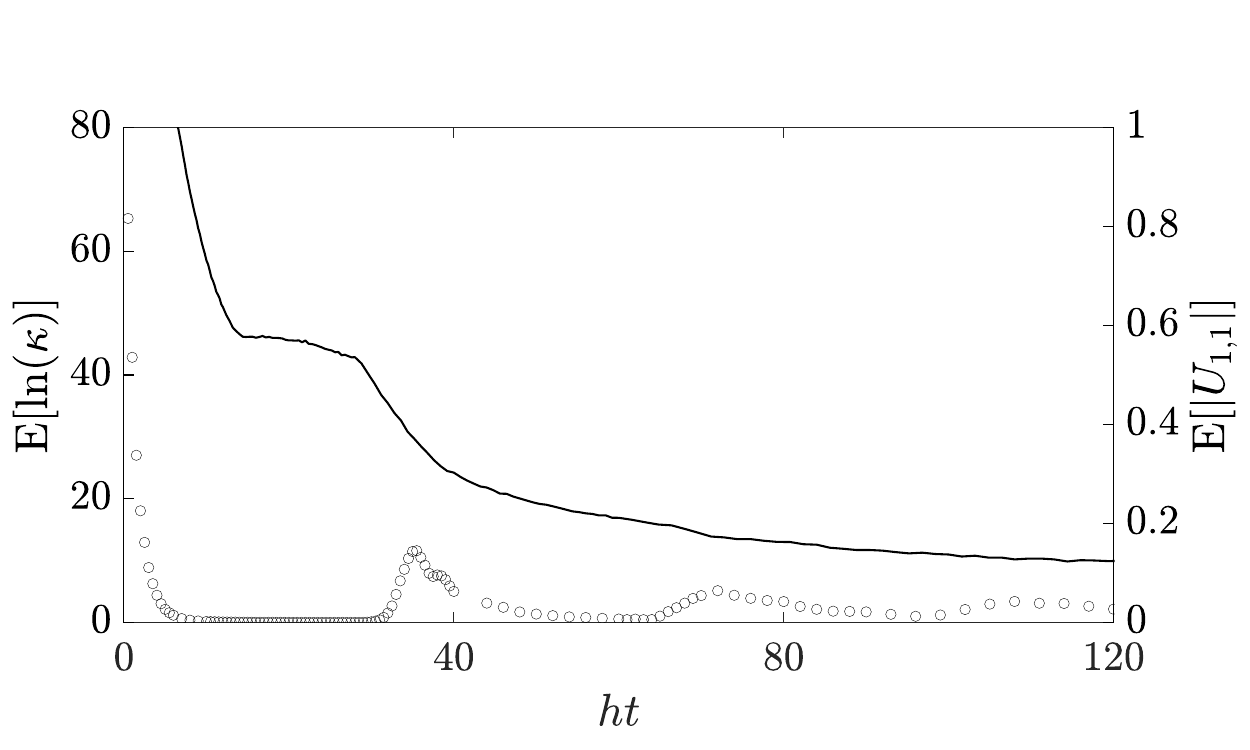}\vspace{-3ex}
	 	\flushleft\normalsize{(c)}\hspace{-2.4ex}\includegraphics[width=\linewidth]{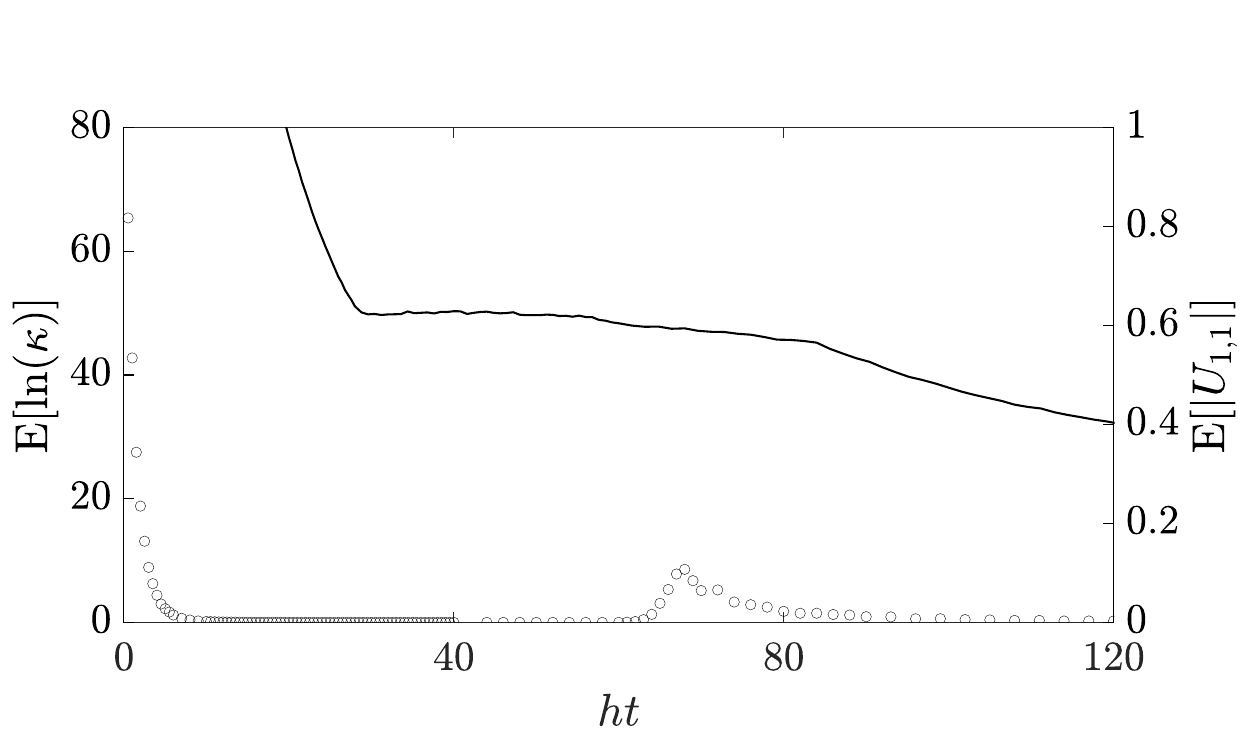}
	 	\caption{\label{fig:U11}The mean of the logarithm of the condition number (solid lines) and the average of the magnitude $|U_{1,1}|$ over 50 realizations (circles) for the  randomly-controlled multilevel system from case study 1. The parameters are the same as in Fig.~\ref{fig:cond_Nlevel}. Here, (a), (b), and (c) respectively depict the dimensions 16, 32, and 64.}
	 \end{figure}

	 In Fig.~\ref{fig:U11}, 
	 	 	 				 we show the evolution of this quantity (circles) for the same parameters as in Fig.~\ref{fig:cond_Nlevel} and compare it with the evolution of $\mathrm{E}[\ln(\kappa)]$ (solid lines). The numerical averages over $50$ realizations are shown in the subfigures (a), (b), and (c) for the three values $16$, $32$, and $64$ of the dimension $d$, respectively. Here, for numerical convenience, in the simulation we made the approximation $\exp[-i(H_0+fH_c)\Delta t]\approx \exp(-iH_0\Delta t)\exp(-ifH_c\Delta t)$, which we checked to be quite accurate for our choice of parameters. 
	 
	 From the numerical findings we conjecture that changes in the slope of $\mathrm{E}[\ln(\kappa)]$ occur around times when $\mathrm{E}[|U_{1,1}|]$ shows revivals. The mechanism of these revivals can be explained by realizing that $|U_{1,1}|=|\langle 1 \vert U(t)\vert 1 \rangle|$, which is nothing but the square root of the population of the state $\vert 1\rangle$, if the system is initially prepared in $\vert 1\rangle$. The chain-like structure of the quantum system suggests that the hopping term transports the population toward the opposite end of the chain, where it is then reflected. This means it takes a certain amount of time until $U_{1,1}$ changes again after its initial drop. This is similarly the case for other entries of $U$. We therefore suggest that the particular way in which changes spread through the system leads to the emergence of the plateaus, which is strongly connected to the coupling graph of the system.

\end{document}